\documentclass[journal]{IEEEtran}
\usepackage{amsmath,amsfonts}
\usepackage{algorithmic}
\usepackage{algorithm}
\usepackage{array}
\usepackage[caption=false,font=normalsize,labelfont=sf,textfont=sf]{subfig}
\usepackage{textcomp}
\usepackage{stfloats}
\usepackage{url}
\usepackage{verbatim}
\usepackage{graphicx}
\usepackage{cite}
\usepackage{pifont}
\usepackage{multirow, amsfonts,bbding, url}
\usepackage{booktabs}
\usepackage{orcidlink}
\usepackage{tabularx}
\begin{document}

\title{Adapting Speech Foundation Models for Unified Multimodal Speech Recognition with Large Language Models}
\author{
Jing-Xuan Zhang\orcidlink{0000-0003-4341-3174}, Genshun Wan\orcidlink{0000-0002-5813-9430}, Jin Li\orcidlink{0000-0002-0260-3169}, Jianqing Gao\orcidlink{0000-0001-5575-4940}, Duo Zhao\orcidlink{0009-0009-8079-3226}, \\ Zhen-Hua Ling\orcidlink{0000-0001-7853-5273},~\IEEEmembership{Senior Member, ~IEEE} %~\IEEEmembership{Staff,~IEEE,} % <-this % stops a space
%\thanks{This work was supported by the National Natural Science Foundation of
%China under Grant No.62401348 and No. 62206162, Fundamental Research
%Funds for the Central Universities under Grant No. GK202406005 \emph{(Corresponding author: Jing-Xuan Zhang and Jianqing Gao)}}
%\thanks{Jing-Xuan Zhang, Jin Li, Duo Zhao are with the School of Artificial Intelligence and
%Computer Science, Shaanxi Normal University (e-mail: \{jxzhanggg, jin.li, zhaoduo\}@snnu.edu.cn). Genshun Wan and Zhen-Hua Ling are with the University of
%Science and Technology of China (e-mail: gswan@mail.ustc.edu.cn, zhling@ustc.edu.cn). Jianqing Gao is with the iFLYTEK Research, iFLYTEK Co., Ltd. (e-mail: jqgao@iflytek.com)}
}

\markboth{Journal of \LaTeX\ Class Files,~Vol.~14, No.~8, August~2021}%
{Shell \MakeLowercase{\textit{et al.}}: A Sample Article Using IEEEtran.cls for IEEE Journals}

\maketitle

\begin{abstract}
While speech foundation models (SFMs) have demonstrated remarkable performance in audio-only tasks, their adaptation to multimodal scenarios remains underexplored. 
This work presents UASR-LLM, a novel framework that adapts frozen SFMs to unified visual speech recognition (VSR), automatic speech recognition (ASR), and audio-visual speech recognition (AVSR) by leveraging large language models (LLMs) as text decoders. Visual representations are injected into multiple SFM layers via visual injection modules, enabling multimodal fusion and unified representation learning. The augmented SFMs are connected to decoder-only LLMs through a feed-forward adaptor, where concatenated representations and instruction prompts guide transcription.
We propose a two-stage training strategy consisting of visual injection pretraining followed by speech recognition finetuning. The pretraining stage aligns audio, visual, and audio-visual representations within the frozen SFM backbone, while the finetuning stage integrates LLMs for unified optimization across speech recognition tasks. 
%During training, only the newly introduced visual injection modules, adaptors, and low-rank adaptation (LoRA) parameters are optimized.
Experimental results demonstrate superior performance over state-of-the-art baselines across VSR, ASR, and AVSR under both clean and noisy conditions. Ablation studies further confirm generalization across various SFMs and LLMs, validating the effectiveness of the proposed training strategy.
\end{abstract}


\section{Introduction}
%\IEEEPARstart{S}{peech} recognition has been greatly advanced by speech foundation models (SFMs) such as WavLM~\cite{chen2022wavlm} and Whisper~\cite{radford2023robust}, which achieve strong automatic speech recognition (ASR) performance through large-scale pretraining. 
\IEEEPARstart{S}{peech} recognition~\cite{9174746,11021228,JMLR:v25:23-1318} has advanced substantially with the emergence of speech foundation models (SFMs), such as WavLM~\cite{chen2022wavlm} and Whisper~\cite{radford2023robust}, which achieve strong performance through large-scale pretraining.
Nevertheless, recognition accuracy remains vulnerable to environmental noise. To improve robustness, visual speech recognition (VSR)~\cite{9837888, 10682067, ZHANG2025126741} and audio-visual speech recognition (AVSR)~\cite{9755926, afouras2018deep, hong2022visual} have been widely explored, exploiting lip movements either as an alternative or as a complement to audio signals. By integrating visual and acoustic cues, AVSR systems substantially enhance noise robustness and have shown increasing practical value in scenarios such as online conferencing and embodied agents.
 
Modern AVSR systems predominantly adopt sequence-to-sequence architectures, including recurrent neural network transducers (RNN-T)~\cite{rouditchenko24_interspeech} and hybrid CTC–attention frameworks~\cite{petridis2018audio, ma2021end, 9755926}. In parallel, audio-visual self-supervised learning~\cite{9837888, shiavhubert, haliassos2024braven} has emerged as an effective pretraining paradigm for leveraging large-scale unlabeled data, with representative models such as AV-HuBERT~\cite{shiavhubert, shi22_interspeech}, REVAn~\cite{haliassos2022jointly, haliassos2024braven}, and AV2vec~\cite{haliassos2022jointly}. After finetuning on limited labeled data, these models achieve strong performance on both VSR and AVSR tasks.
Building on these advances, Haliassos et al. proposed unified speech recognition (USR)~\cite{haliassos2024unified}, which jointly supports ASR, VSR, and AVSR within a single model, substantially reducing training and deployment costs compared to modality-specific systems. More recently, speech foundation models (SFMs) have been explored for VSR~\cite{djilali2023lip2vec, prajwal24_interspeech} and AVSR~\cite{han-etal-2024-xlavs, rouditchenko24_interspeech}. Zhang et al.~\cite{ZHANG2025111432} further demonstrated that SFMs can serve as effective teachers for knowledge distillation in audio-visual self-supervised pretraining, achieving competitive results across ASR, VSR, and AVSR benchmarks. In addition, several studies have investigated the direct adaptation of SFMs to audio-visual settings, such as Lip2vec~\cite{djilali2023lip2vec} and Whisper-Flamingo~\cite{baevski2020wav2vec, 10999072}.

Despite recent progress, most existing approaches still treat ASR, VSR, and AVSR as separate tasks with modality-specific architectures. 
While USR~\cite{haliassos2024unified} builds a unified model from scratch, we instead adapt existing large-scale pretrained models for unified multimodal speech recognition in a parameter-efficient manner.
To this end, we propose UASR-LLM (\textbf{U}nified \textbf{A}daptive \textbf{S}peech \textbf{R}ecognition with \textbf{L}arge \textbf{L}anguage \textbf{M}odels), which adapts frozen speech foundation models for unified ASR, VSR, and AVSR by employing an LLM as a universal text decoder. Specifically, visual representations extracted from a pretrained visual encoder are injected into each SFM block via lightweight visual injection modules, enabling cross-modal integration while preserving the pretrained acoustic structure. The resulting multimodal SFM outputs are projected through a learnable adapter into the hidden embedding space of the LLM, which directly performs text generation across all input modalities, thereby establishing a unified decoding framework.

To enable SFMs to process visual and audio-visual inputs, we introduce a visual injection pretraining stage based on cross-modal knowledge distillation. This is followed by a unified speech recognition finetuning stage, where instruction-guided random modality dropout is applied to enhance robustness. Throughout training, both the SFM and visual encoder remain frozen, while the LLM is efficiently finetuned using low-rank adaptation (LoRA)~\cite{hu2022lora}. This design substantially reduces memory and computational costs while preserving the generalization ability of pretrained models.
By using a single SFM as a shared multimodal backbone, our framework facilitates knowledge transfer from audio to visual inputs and aligns representations across modalities, enabling the LLM to generate transcriptions in a unified output space and seamlessly perform ASR, VSR, and AVSR within one model.

Extensive experiments demonstrate that UASR-LLM consistently outperforms state-of-the-art baselines across ASR, VSR, and AVSR when trained with comparable amounts of audio-visual data. On the LRS3 benchmark, our model achieves word error rates of 20.9\%, 0.84\%, and 0.69\% for VSR, ASR, and AVSR, respectively. Notably, under noisy conditions, UASR-LLM exhibits strong robustness, yielding substantial WER reductions over competing methods. We further show that the proposed framework generalizes well across different choices of speech foundation models and large language models, with all configurations surpassing baseline performance. Finally, ablation studies validate the effectiveness of our training strategy and confirm the advantages of unified speech recognition within a single model architecture.

\section{Related works}

\subsection{Speech foundation models}

Recent years have witnessed the emergence of various speech foundation models, which can be broadly categorized into two paradigms: self-supervised~\cite{baevski2020wav2vec,hsu2021hubert,baevski2022data2vec} and supervised approaches~\cite{radford2023robust, peng25c_interspeech}.
Among SSL-based speech foundation models, 
HuBERT~\cite{hsu2021hubert}  utilizes k-means clustering to generate pseudo-class labels for model training using a mask prediction loss,
with iterative refinement through subsequent clustering and mask prediction steps.
WavLM~\cite{hsu2021hubert} improves HuBERT by using a more diverse pretraining dataset and performs speech denoising modeling during pretraining.
On the other hand, Whisper family~\cite{radford2023robust} exemplifies supervised learning-based speech foundation model. Trained on large-scale multilingual and multitask labeled data, it demonstrates robust performance in ASR and speech translation tasks.

\subsection{Visual representation learning}
Visual representation learning is crucial for the performance of VSR and AVSR tasks, which can be broadly categorized into supervised and self-supervised approaches.
Supervised methods leverage labeled~\cite{Chung_2017_CVPR} or weakly labeled audio-visual speech data~\cite{10096889} to optimize representation extraction and text decoding in an end-to-end manner through label prediction. Several auxiliary tasks have been proposed to enhance visual encoder learning, including articulation feature prediction~\cite{sterpu2020teach} and distillation from acoustic models~\cite{sterpu2020teach, ma2022training, afouras2020asr}.
Self-supervised methods utilize unlabeled audio-visual data with pretext tasks such as masked prediction loss functions. Representative approaches include AV-HuBERT~\cite{shiavhubert,hsu2022u}, RAVEn~\cite{haliassos2022jointly, haliassos2024braven}, AV2vec~\cite{Zhang2023av2vec}, and AV-data2vec~\cite{Lian2023avdata2vec}. Zhang et al.~\cite{ZHANG2025111432} proposed a joint audio-visual representation learning method that leverages cross-modal knowledge distillation from pretrained SFMs.
%This work adopts our previously pretrained model for visual representation extraction, which can be viewed as an extension of our previous study. 
Here, we further leverage SFMs directly for processing audio-visual data, which can be viewed as an extension of this study.
%moving beyond the distillation approach to direct utilization of foundation models.

\subsection{Speech recognition with LLMs}

Large language models (LLMs) have attracted significant attention for speech recognition applications, with research broadly falling into two categories: multimodal LLMs for universal AI assistants~\cite{zhang-etal-2023-speechgpt, Qwen2.5-Omni} and LLMs for enhancing traditional speech recognition tasks~\cite{10389703,10447605,10445874}. Our work belongs to the latter category.
Recent approaches for integrating LLMs with speech recognition typically employ lightweight adapters to connect frozen speech encoders with LLMs. Representative works include joint speech-language models with small adapters~\cite{10389703} and Q-Former-based connections~\cite{10445874}. Some studies explore discrete speech representations as LLM inputs~\cite{zhang-etal-2023-speechgpt, xu24d_interspeech}.
LLMs have also been adopted for visual and audio-visual speech recognition~\cite{yeo-etal-2024-visual, yang24f_interspeech,10889251}. VSP-LLM~\cite{yeo-etal-2024-visual} focuses on visual speech recognition, while Llama-AVSR~\cite{10889251} uses separate audio and visual encoders with temporal concatenation as LLM input. MMS-Llama~\cite{yeo2025mms} employs audio-visual fusion within encoders and Q-former adapters for improved efficiency.
Our work differs by proposing a unified multimodal speech recognition approach that processes both audio and visual modalities within a single model architecture, rather than targeting specific modalities separately.

\section{Proposed method}

\begin{figure*}
    \centering
    \includegraphics[width=0.7\textwidth]{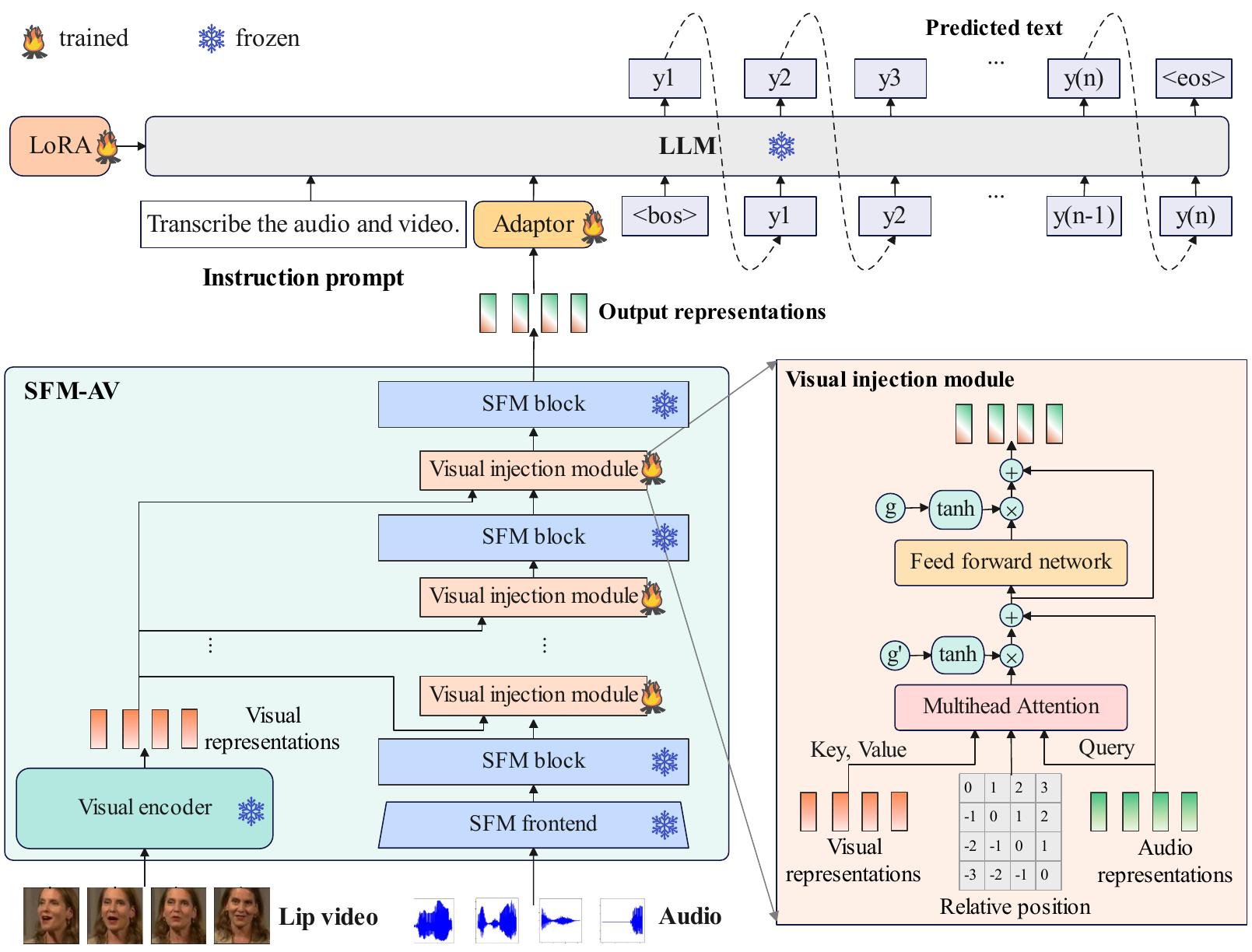}
    \caption{Overall architecture of our proposed method when performing audio-visual speech recognition.}
    \label{fig:fig1}
\end{figure*}

Our UASR-LLM method comprises two main components: an audio-visual adapted speech foundation model (SFM-AV) 
that processes both audio and visual modalities and encodes them into a unified representational space, 
and a large language model that decodes the encoded representations into text. 
The SFM-AV consists of a visual encoder and a speech foundation model enhanced with visual injection modules. 
Our model architecture is illustrated in Figure~\ref{fig:fig1}.
The training of our model follows a two-stage process: in the first pretraining stage, 
the SFM is adapted to encode both audio and visual information into a unified hidden space; 
in the second finetuning stage, the adapted SFM is connected to the LLM to enable unified multimodal speech recognition. 
The detailed model architecture and training strategy are described in the following sections.

\subsection{Model structure}

\subsubsection{Visual encoder}

Let lip video be $\mathbf{V} \in \mathbb{R}^{T_v \times H \times W \times C}$, where $T_v, H, W,$ and $C$ represent the temporal, height, width, and channel dimensions, respectively. The visual encoder $\text{E}_v$ processes the visual input into hidden representations:
$\mathbf{H}_v = \text{E}_v(\mathbf{V}) \in \mathbb{R}^{T_v \times D_v}$, 
where $D_v$ is the feature dimension of the encoded representations.
In this work, we adopt the visual component of the pretrained DistillAV model~\cite{ZHANG2025111432} as our visual encoder. DistillAV is designed to distill knowledge from speech foundation models (SFMs) for joint audio-visual representation learning. Unlike SFMs, which are trained on large-scale speech-only corpora, DistillAV is trained on relatively smaller audio-visual datasets. As a result, its performance in audio-only tasks is inferior to that of its teacher models, as shown in previous ASR evaluations~\cite{ZHANG2025111432}. However, DistillAV exhibits strong visual modeling capabilities and consistently outperforms state-of-the-art lipreading models such as AV-HuBERT~\cite{shiavhubert} and AV2vec~\cite{Zhang2023av2vec}. Therefore, we use the pretrained DistillAV model to extract representations from lip video inputs.

\subsubsection{Visual injection module}
The visual injection module is introduced to incorporate visual information into the pretrained SFM. For the $i$-th block of the SFM, we employ the audio representations $\mathbf{H}_a^i \in \mathbb{R}^{T \times D}$ as the query and attend to visual representations through a cross-attention mechanism. To enable the model to exploit relative position information between the two modalities, we incorporate relative position embeddings~\cite{shaw-etal-2018-self}.
The output is subsequently processed through a feed-forward network and then fed into the next SFM block. Both the cross-attention and feed-forward network are augmented with residual connections and tanh gating, following the approach in Whisper-Flamingo~\cite{rouditchenko24_interspeech}. 

Formally, our visual injection module operates as follows: 
\begin{equation} 
\mathbf{H}^i_{mid} = \text{tanh}(g_i) \odot \text{MHA}(\mathbf{H}_a^i, \mathbf{H}_v, \mathbf{H}_v, \mathbf{D}_{av}) + H_a^i, 
\end{equation} 
\begin{equation} 
\mathbf{H}^i_{av} = \text{tanh}(g'_i) \odot \text{FFN}(\mathbf{H}^i_{mid}) + \mathbf{H}^i_{mid}, 
\end{equation} 
where $\text{MHA}()$ and $\text{FFN}()$ represent the multi-head attention and feed-forward network modules, respectively. $\mathbf{D}_{av} = [d_{m,n}] \in \mathbb{R}^{T \times T_v}$, where $d_{m, n}$ represents the relative position encoding between the $m$-th audio frame and the $n$-th video frame. $g_i$ and $g_i'$ are trainable gating parameters, which are initialized to zero.
It provides an initialization state identical to the original SFMs, which helps stabilize the training.
The visual injection modules are inserted before each block of the SFM, while the SFM parameters remain frozen throughout training.

\subsubsection{Prompting the LLM for USR}

The output representations from the SFM-AV are transformed through a two-layer feed-forward adaptor to match the embedding dimension of large language models, serving as continuous audio-visual tokens for prompting the LLM to generate transcribed text. We employ task-specific instructions that prepend the audio-visual tokens according to the unified speech recognition task. The instruction prompt follows the template ``Transcribe the [].'', where the placeholder [] is filled with ``audio'', ``video'', or ``audio and video'' for ASR, VSR, and AVSR tasks, respectively. 
For efficient LLM finetuning, we adopt the low-rank adaptation (LoRA) method. The low-rank parameters are applied to adapt all blocks' self-attention components as well as the output projection layer for predicting text logits. Unlike previous approaches such as Llama-AVSR's frame-compression~\cite{10889251} or MMS-Llama's Q-former~\cite{yeo2025mms}, our method does not employ frame reduction techniques. This design choice reflects our primary focus on improving recognition performance rather than computational efficiency, with the latter left for future investigation.

\begin{figure*}
    \centering
    \includegraphics[width=0.7\textwidth]{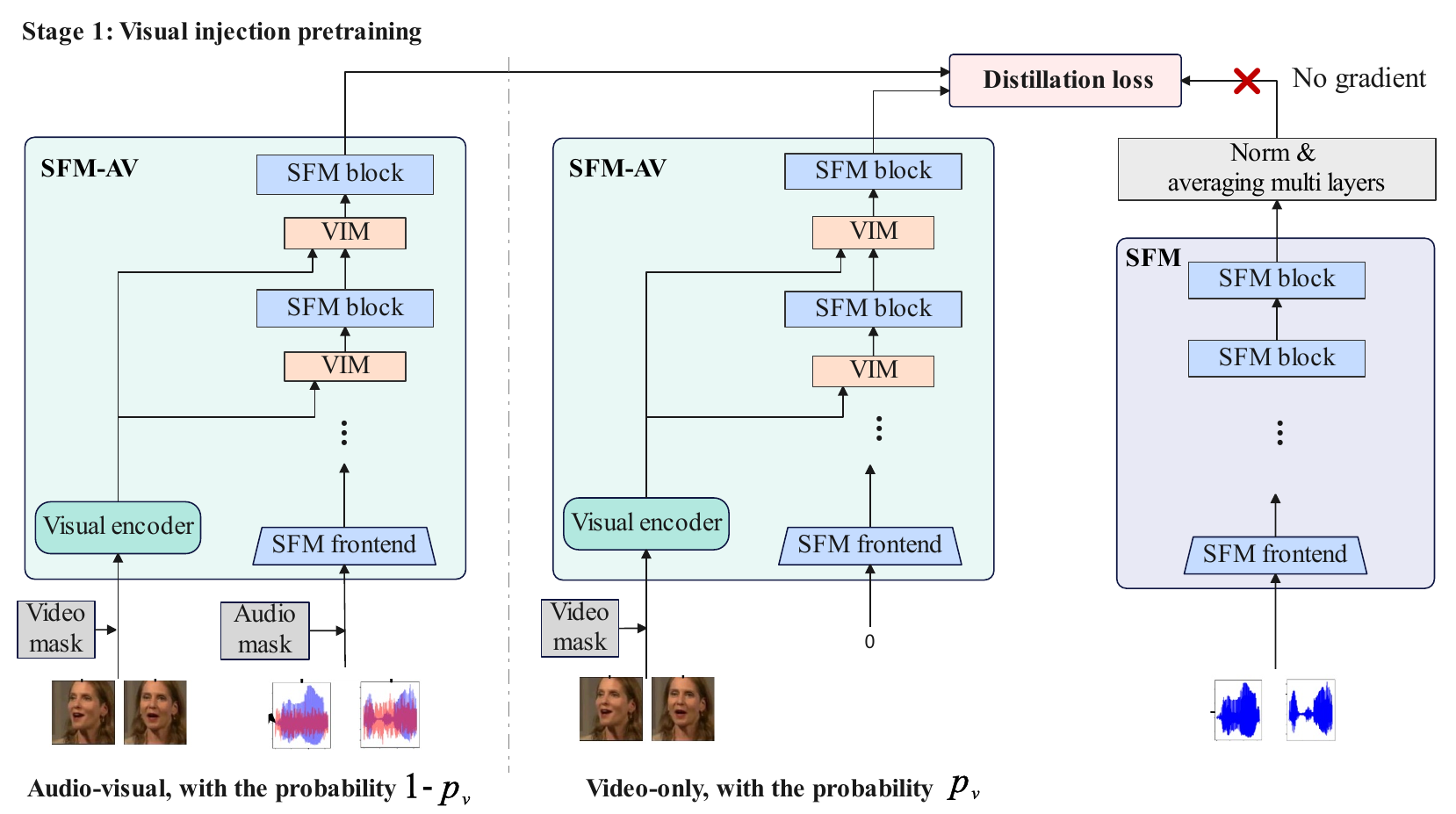}
    \caption{Illustratation of the visual injection pretraining stage of our proposed method.}
    \label{fig:fig2}
\end{figure*}

\subsection{Training strategy}
\subsubsection{Visual injection pretraining}
The first stage of our pretraining strategy focuses on integrating visual information into the speech foundation model to enable SFM-AV to process multimodal inputs and generate unified audio-visual representations. We employ a cross-modal knowledge distillation approach where clean audio representations from the original SFM serve as supervision signals for training SFM-AV with corrupted audio-visual inputs, as shown in Figure~\ref{fig:fig2}.

Specifically, we extract latent representations from the last $k$ blocks of the SFM when processing clean audio $\mathbf{A}$:
\begin{equation}
{\mathbf{H}^{L-k+1}, \dots, \mathbf{H}^{L}} = \text{SFM}(\mathbf{A}),
\end{equation}
where $L$ denotes the total number of blocks in the SFM. To obtain the final teacher representation $H^{\mathcal{T}}$, we apply instance normalization to each block's output and compute their average:
\begin{equation}
\label{eq:eq2}
\mathbf{H}^{\mathcal{T}} = \frac{1}{k} \sum_{i=L-k+1}^{L} \text{IN}(\mathbf{H}^i),
\end{equation}
where $\text{IN}(\cdot)$ represents instance normalization. This multi-layer aggregation captures rich hierarchical features, which follows previous study~\cite{ZHANG2025111432}.

For SFM-AV pretraining, we apply several data corruption strategies to promote effective audio-visual fusion when predicting clean audio representations. First, the input audio is corrupted by randomly mixing it with background noise, resulting in $\mathbf{A}^N$. Both the noisy audio $\mathbf{A}^N$ and the video $\mathbf{V}$ are then corrupted using span-based masking. Specifically, for each modality, we randomly select a starting frame and replace a consecutive span of frames with zeros starting from that point.
In addition, we adopt modality dropout during training. With probability $p_v$, only the video input is provided to the model, while the audio input is replaced with zeros. With probability $1 - p_v$, both audio and video inputs are used.
After applying noise mixing, span masking, and modality dropout, SFM-AV processes the corrupted inputs to generate output representations $\mathbf{H}^o$, which is summarized as:
\begin{equation}
\mathbf{H}^o = \left \{
\begin{array}{l l}
\text{SFM-AV}(0, \mathcal{M}_v(\mathbf{V})) & \text{with }~p_v, \\
\text{SFM-AV}(\mathcal{M}_a(\mathbf{A}^N), \mathcal{M}_v(\mathbf{V})) & \text{with }~1 - p_v,
\end{array}
\right.
\end{equation}
where $\mathcal{M}_a$ and $\mathcal{M}_v$ denote the span masking functions applied to the audio and video modalities, respectively.

The training objective employs a knowledge distillation loss that combines L1 distance and negative cosine similarity:
\begin{equation}
L_{dis} = \frac{1}{T} \sum_{i=1}^{T} \left(
|| \mathbf{H}^{o}_i W - \mathbf{H}^{\mathcal{T}}_i ||_1 - \cos(\mathbf{H}^{o}_i W, \mathbf{H}^{\mathcal{T}}_i)
\right),
\end{equation}
where $W$ represents learnable projection parameters.

\begin{figure*}
    \centering
    \includegraphics[width=0.7\textwidth]{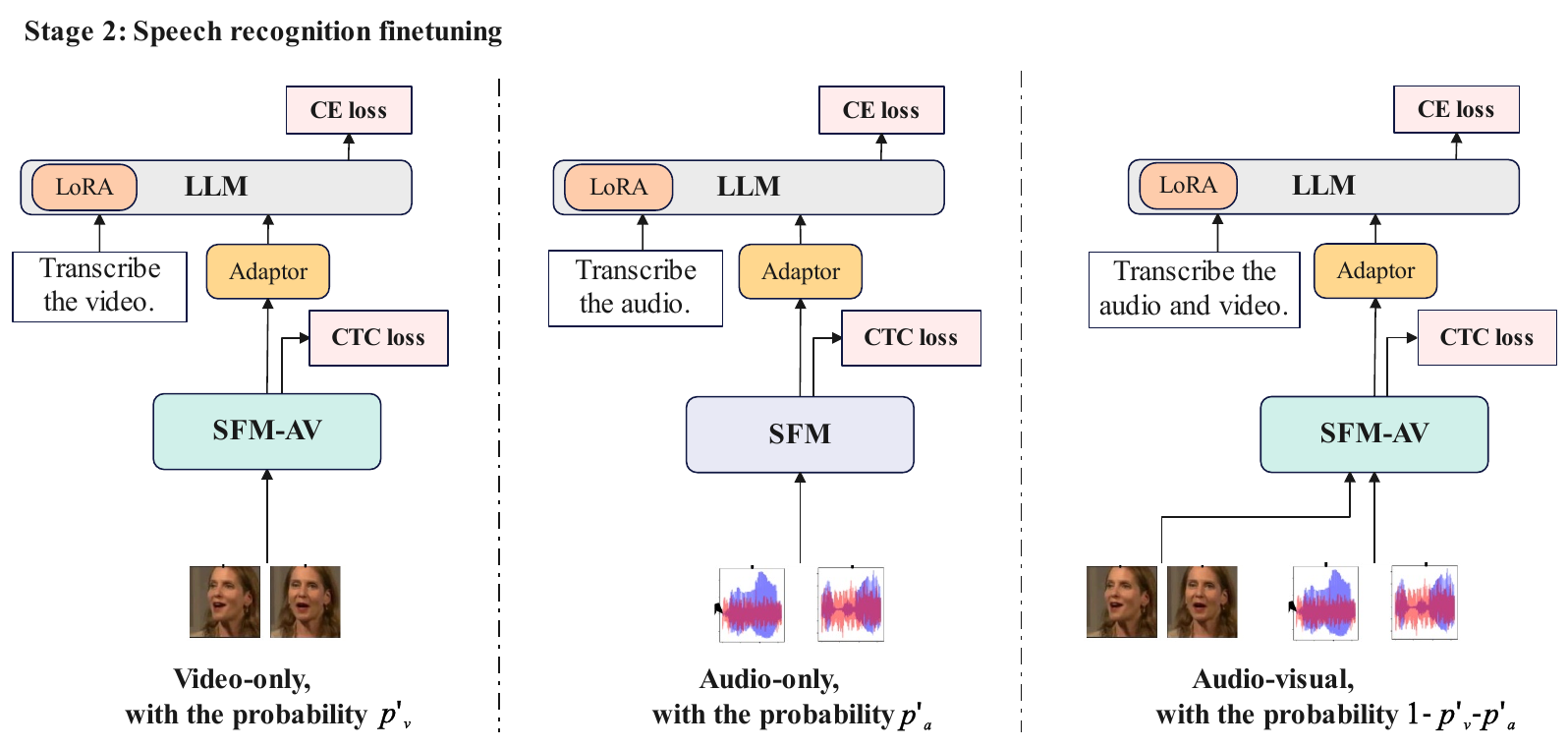}
    \caption{Illustration of the speech recognition finetuning stage of our proposed method.}
    \label{fig:fig3}
\end{figure*}

\subsubsection{Speech recognition finetuning}
Following the pretraining stage, we connect the SFM-AV encoder with an LLM to perform unified speech recognition across multiple modalities,
as shown in Figure~\ref{fig:fig3}. The model pathway varies depending on the target task: for VSR and AVSR, we utilize SFM representations enhanced with visual information through the injection module. For ASR, we bypass the visual injection module and employ the original SFM to process audio-only inputs.
To enhance model robustness, we employ a modality dropout strategy with probabilities $p_v'$, $p_a'$, and $1-p_a'-p_v'$ for visual-only, audio-only, and audio-visual combined inputs, respectively. The instruction prompts are selected accordingly to match the input modality configuration.

The primary training loss is cross-entropy $L_{CE}$ for text prediction by the LLM. Additionally, we incorporate an auxiliary CTC loss to optimize the encoder directly. This requires connecting the encoder to an additional feed-forward projection layer for logit prediction. Initial experiments used the same vocabulary as the LLM; however, convergence proved challenging due to the large vocabulary size (e.g., ~152k tokens for Qwen 2.5). Therefore, we adopted a separate, smaller 1k vocabulary specifically for the CTC loss, which significantly improved training stability.
The total loss combines both objectives:
\begin{equation}
\label{eq:eq1}
L = L_{CE} + \lambda L_{CTC},
\end{equation}
where $\lambda$ controls the relative importance of the auxiliary CTC loss.
During finetuning, we jointly optimize the LoRA parameters of the LLM, the feed-forward adaptor, and the visual injection module, while keeping the core SFM parameters frozen.

\section{Experiments and results}

\subsection{Implementation details}
\label{sec:implement}
\textbf{Datasets.} 
For pretraining, we utilized the LRS3 dataset\footnote{\url{https://mmai.io/datasets/lip_reading/}}, which comprises approximately 433 hours of English audio-visual speech data. We also employed a combination dataset of LRS3 and the English-only version of VoxCeleb2\footnote{\url{https://www.robots.ox.ac.uk/~vgg/data/voxceleb/vox2.html}}, curated by Shi et al.~\cite{shiavhubert}, totaling approximately 1759 hours of audio-visual speech data. The VoxCeleb data was transcribed using the Whisper-large-v2 model\footnote{\url{https://huggingface.co/openai/whisper-large-v2}}.
For finetuning, we adopted both a 30-hour subset and the full 433-hour training set of LRS3. Additionally, the model can be finetuned on the complete 1759-hour audio-visual speech dataset. A validation set of 1000 utterances was reserved, and results were reported on the LRS3 test set. We used the facial images provided in the original datasets without cropping to the lip region-of-interest (ROI)~\cite{Zhang2022face}.

\textbf{Noise augmentation.} 
The MUSAN dataset\footnote{\url{https://www.openslr.org/17/}}  was employed for noise augmentation. Following the protocol outlined in~\cite{DBLP:journals/corr/abs-2201-01763}, we sampled and added noise to the audio input of the student model during training. To evaluate our proposed method for AVSR under noisy conditions, we constructed test sets with various noise types, including music, natural sounds, speech, and babble, at signal-to-noise ratios (SNRs) of -10, -5, 0, 5, and 10 dB.

\textbf{Model configurations.}
Our experiments employed a default configuration consisting of three main components: the DistillAV model~\cite{ZHANG2025111432} as the visual encoder, WavLM-large~\cite{chen2022wavlm} (317M parameters) as the speech foundation model (SFM), and Qwen 2.5-7B\footnote{\url{https://huggingface.co/Qwen/Qwen2.5-7B}} as the large language model. The visual encoder offered two variants: Base (103M parameters) and Large (325M parameters). Following previous studies~\cite{shiavhubert,ZHANG2025111432}, we denote the proposed method as UASR-LLM-B and UASR-LLM-L accordingly.
The visual injection modules were configured with 8 heads, 512 hidden dimensions, and 256 dimensions for both cross-attention and feed-forward networks, totaling approximately 58M parameters. The two-layer feed-forward network adapter that connects the SFM and LLMs contains around 10M parameters. LoRA weights were configured with a rank of 16, resulting in 12M total parameters.
To demonstrate the generalization capability of our method, we explored additional speech foundation models: the open-source Whisper-medium~\cite{radford2023robust} and an internal Conformer-based ASR model (Conformer-Speech), both containing approximately 300M parameters comparable to WavLM-large. We also evaluated ChatGLM v3-6B\footnote{\url{https://huggingface.co/THUDM/chatglm3-6b}} as an alternative large language model.

The WavLM model generates representations at 50 fps, double the frame rate of visual representations. We set relative positional encoding according to the real-time interval between audio and visual frames using sinusoidal positional embedding. The parameter $k$ in Equation~\ref{eq:eq2} was set to 8 for WavLM-based SFM and 1 for Whisper and Conformer-Speech SFM, following the previous study~\cite{ZHANG2025111432}.

During pretraining, 80\% of audio features and 30\% of video features were masked for the student model, with modality dropout probability $p_v$ set to 0.5. We employed the Adam optimizer with a learning rate schedule that linearly warmed up to 0.0005 over the initial 5\% of updates, maintained this rate for 85\% of updates, then exponentially decayed over the remaining 10\%. The total number of updates was 100k and 200k for 433h and 1759h pretraining data, respectively. During finetuning, modality dropout probabilities $p'_v$ and $p'_a$ were set to 0.5 and 0.25, respectively. For the auxiliary CTC loss, we utilized subword units with a vocabulary size of 1000 as targets, with $\lambda$ in Equation~\ref{eq:eq1} set to 0.25. The learning rate warmed up to 0.0002 over the initial one-third of updates, followed by exponential decay. The total number of updates was 30k, 60k, and 120k for 30h, 433h, and 1759h finetuning data, respectively.

\subsection{Comparison with state-of-the-art baselines}

\begin{table*}[]
    \centering
     \caption{ Word error rate (WER) (\%) on LRS3 test set using low-resource (30h) labeled audio-visual data. "Shared" indicates whether the same model is used across ASR, VSR, and AVSR tasks.}
    \label{tab:tab1}
    \scalebox{1.0}{
    \begin{tabular}{c c c c c c c c c}
     \toprule
    Methods & Unlab. & Lab. & Shared & VSR & ASR & AVSR & ASR (Noisy) & AVSR (Noisy) \\
    \midrule
    AV-HuBERT-B~\cite{shiavhubert} & \multirow{11}*{433h} & \multirow{11}*{30h} & \ding{55} & 51.8 & 4.9 & 4.7 & -- & --\\
    AV-HuBERT-L~\cite{shiavhubert} &  &  & \ding{55} & 44.8 & 4.5 & 4.2 & -- & -- \\
    VATLM-B~\cite{zhu2023vatlm} & &  & \ding{55} & 48.0 & 3.6 & -- & -- \\
    RAVEn-B~\cite{haliassos2022jointly} & &  & \ding{55} &  47.0 & 4.7 & --  & -- \\
    BRAVEn-B~\cite{haliassos2024braven} & & & \ding{55} & 43.4 & 4.0 & --  & -- & -- \\
    AV-data2vec-B~\cite{Lian2023avdata2vec} & & & \ding{55} & 45.2 & 4.4 & 4.2 & -- & -- \\
    AV2vec-MLM~\cite{Zhang2023av2vec} & & & \ding{55} & 39.4 & 5.6 & 5.4 & -- & -- \\
    USR~\cite{haliassos2024unified} & & & \ding{51} & 36.0 & 3.2 & 3.0 & -- & -- \\
    DistillAV-B~\cite{ZHANG2025111432} & & & \ding{55} & 37.2 & 3.0 & 2.8 & 50.5 & 11.4 \\
    DistillAV-L~\cite{ZHANG2025111432} & & & \ding{55} & 33.2 & 2.4 & \textbf{1.8} & 45.3 & 11.9 \\
    \emph{UASR-LLM-B (Ours)} & & & \ding{51} & 31.7 & \textbf{2.2} & 2.1 & 39.2 & 4.8 \\
    \emph{UASR-LLM-L (Ours)} & & & \ding{51} & \textbf{30.6} & \textbf{2.2} & 2.0 & \textbf{38.7} & \textbf{4.4} \\
    \midrule
    AV-HuBERT-B~\cite{shiavhubert}  & \multirow{16}*{1759h} & \multirow{16}*{30h} & \ding{55} & 46.1 & 4.6 & 4.0 & -- & -- \\
    AV-HuBERT-L~\cite{shiavhubert}  & & & \ding{55} & 32.5 & 2.9 & 3.3 & -- & -- \\
    VATLM-B~\cite{zhu2023vatlm}  & & & \ding{55} & 42.6 & -- & 3.4 & -- & -- \\
    VATLM-L~\cite{zhu2023vatlm}  & & & \ding{55} & 31.6 & -- & 2.7 & -- & -- \\
    BRAVEn-B~\cite{haliassos2024braven} & & & \ding{55} & 35.1 & 3.0 & -- & -- & -- \\
    BRAVEn-L~\cite{haliassos2024braven} & & & \ding{55} & 30.8 & 2.3 & -- & -- & -- \\
    AV-data2vec-B~\cite{Lian2023avdata2vec} & & & \ding{55} & 37.8 & 3.7 & 3.3 & -- & -- \\
    AV-data2vec-L~\cite{Lian2023avdata2vec} & & & \ding{55} & 30.8 & 2.7 & 2.7 & -- & -- \\
    USR-B~\cite{haliassos2024unified} & & & \ding{51} & 28.4 & 2.6 & 2.5 & -- & -- \\
    USR-L~\cite{haliassos2024unified} & & & \ding{51} & 26.9 & 2.4 & 2.4 & -- & -- \\
    DistillAV-B~\cite{ZHANG2025111432} & & & \ding{55} & 36.9 & 2.5 & 2.2 & 44.6 & 9.2 \\
    DistillAV-L~\cite{ZHANG2025111432} & & & \ding{55} & 30.2 & 2.2 & 2.1 & 41.4 & 8.1 \\
    LLama-AVSR~\cite{10889251} & & & \ding{55} & 28.4 & \textbf{1.5} & -- & -- & -- \\
    \emph{UASR-LLM-B (Ours)} & & & \ding{51} & 27.3 & 2.3 & \textbf{1.8} & \textbf{38.3} & \textbf{3.5} \\
    \emph{UASR-LLM-B* (Ours)} & & & \ding{51} & 28.4 & 2.4 & 1.9 & 39.8 & 3.8 \\
    \emph{UASR-LLM-L (Ours)} & & & \ding{51} & \textbf{25.3} & 2.5 & 1.9 & 39.2 & \textbf{3.5} \\
    \bottomrule
    \multicolumn{9}{l}{
    *Visual encoder pretraining used 1759 hours of data, visual injection pretraining used 433 hours. 
    }
    \end{tabular}
    }
\end{table*}

\begin{table*}[]
    \centering
    \caption{Word error rate (WER) (\%) on LRS3 test set using high-resource (433h or 1759h) labeled audio-visual data.  "Shared" indicates whether the same model is used across ASR, VSR, and AVSR tasks.}
    \label{tab:tab2}
    \scalebox{1.0}{
    \begin{tabular}{c c c c c c c c c}
    \toprule
    Methods & Unlab. & Lab. & Shared & VSR & ASR & AVSR & ASR (Noisy) & AVSR (Noisy) \\
    \midrule
    Makino et al.~\cite{makino2019recurrent} & -- & 31k & \ding{55} & 33.6 & 4.8 & 4.5 & -- & -- \\
    Auto-AVSR~\cite{10096889} & -- & 1.9kh & \ding{55} & 23.5 &  \textbf{1.0} & 1.0 & -- & -- \\
    Auto-AVSR~\cite{10096889} & -- & 3.5kh & \ding{55} & 19.1 & \textbf{1.0} & 
    \textbf{0.9} & -- & -- \\
    LP-Conformer~\cite{10446532} & -- & 100k & \ding{55} & \textbf{12.8} & -- & \textbf{0.9} & -- & -- \\
    \midrule
    %AV-HuBERT-B~\cite{shiavhubert} & \multirow{11}*{433h} & \multirow{11}*{433h} & 44.0 & 3.0 & 2.8 & -- & --\\
    AV-HuBERT-L~\cite{shiavhubert} &  \multirow{9}*{433h} &  \multirow{9}*{433h}  & \ding{55} & 41.6 & 2.7 & 2.5 & -- & -- \\
    BRAVEn-B~\cite{haliassos2024braven} & & & \ding{55} & 36.0 & 1.9 & --  & -- & -- \\
    AV-data2vec-B~\cite{Lian2023avdata2vec} & & & \ding{55} & 39.0 & 2.0 & 1.8 & -- & -- \\
    AV2vec-MLM~\cite{Zhang2023av2vec} & & & \ding{55} & 34.4 & 2.7 & 2.5 & -- & -- \\
    USR~\cite{haliassos2024unified} & & & \ding{51} & 34.3 & 1.9 & 1.6 & -- & -- \\
    DistillAV-B~\cite{ZHANG2025111432} & & & \ding{55} & 34.3 & 1.9 & 1.8 & 44.8 & 9.6 \\
    DistillAV-L~\cite{ZHANG2025111432} & & & \ding{55} & 31.5 & 1.8 & 1.6 & 41.8 & 8.6 \\
    \emph{UASR-LLM-B (Ours)} & & & \ding{51} & 28.6 & 1.1 & 0.93 & 38.1 & 3.4 \\
    \emph{UASR-LLM-L (Ours)} & & & \ding{51} & \textbf{26.4} & \textbf{0.96} & \textbf{0.82} & 
    \textbf{36.8} & \textbf{3.2} \\
    \midrule
    %AV-HuBERT-B~\cite{shiavhubert}  & \multirow{16}*{1759h} & \multirow{16}*{433h} & 34.8 & 2.0 & 1.8 & -- & -- \\
    AV-HuBERT-L~\cite{shiavhubert}  & \multirow{13}*{1759h} &\multirow{13}*{433h} & \ding{55} & 28.6 & 1.3 & 1.4 & -- & -- \\
    %VATLM-B~\cite{zhu2023vatlm}  & & & 34.2 & -- & 1.7 & -- & -- \\
    VATLM-L~\cite{zhu2023vatlm}  & & & \ding{55}  & 28.4 & -- & 1.2 & -- & -- \\
    %BRAVEn-B~\cite{haliassos2024braven} & & & 28.8 & 1.4 & -- & -- & -- \\
    BRAVEn-L~\cite{haliassos2024braven} & & & \ding{55}  & 26.6 & 1.2 & -- & -- & -- \\
    AV-data2vec-B~\cite{Lian2023avdata2vec} & & & \ding{55}  & 32.7 & 1.7 & 1.4 & -- & -- \\
    AV-data2vec-L~\cite{Lian2023avdata2vec} & & & \ding{55}  & 28.5 & 1.4 & 1.3 & -- & -- \\
    USR-B~\cite{haliassos2024unified} & & & \ding{51}  & 26.5 & 1.6 & 1.3 & -- & -- \\
    USR-L~\cite{haliassos2024unified} & & & \ding{51}  & 22.3 & 1.2 & 1.1 & -- & -- \\
    DistillAV-B~\cite{ZHANG2025111432} & & & \ding{55}  & 31.4 & 1.9 & 1.7 & 42.2 & 7.8 \\
    DistillAV-L~\cite{ZHANG2025111432} & & & \ding{55}  & 26.2 & 1.4 & 1.3 & 39.2 & 6.9 \\
    Llama-AVSR~\cite{10889251} & & & \ding{55}  & 25.3 & 1.1 & 0.95 & -- & -- \\
    \emph{UASR-LLM-B (Ours)} & & & \ding{51}  & 23.4 & 0.91 & \textbf{0.73} & 37.8 & \textbf{2.4} \\
    \emph{UASR-LLM-B* (Ours)} & & & \ding{51}  & 23.9 & \textbf{0.86} & 0.78 & 37.0 & 2.6 \\
    \emph{UASR-LLM-L (Ours)} & & & \ding{51}  & \textbf{21.4} & 0.87 & 0.80 & \textbf{35.4} & 2.7 \\
    \midrule
    % AV-HuBERT-L w/ ST~\cite{shiavhubert}
    Llama-AVSR~\cite{10889251} & \multirow{4}*{1759h} & \multirow{4}*{1759h} & \ding{55}  & 24.0 & \textbf{0.81} & 0.77 & -- & -- \\
    RAVEn w/ ST~\cite{haliassos2022jointly} & & & \ding{55}  & 23.1 & 1.4 & -- & -- \\
    USR~\cite{haliassos2024unified} & & & \ding{51}  & 21.5 & 1.2 & 1.1 & -- & -- \\
    \emph{UASR-LLM-L (Ours)} & & & \ding{51}  & \textbf{20.9} & 0.84 & \textbf{0.69} & \textbf{37.2} & \textbf{2.4} \\
    \bottomrule
     \multicolumn{9}{l}{
    *Visual encoder pretraining used 1759 hours of data, visual injection pretraining used 433 hours.
    }
    \end{tabular}
    }
    
\end{table*}

Table~\ref{tab:tab1} and Table~\ref{tab:tab2} present experimental results for models finetuned with low-resource (30h) and high-resource (433h or 1759h) labeled data, respectively. The unlabeled data refers to audio-visual data used for pretraining the DistillAV-based visual encoder and visual injection pretraining, while the labeled data denotes audio-visual data employed for speech recognition finetuning.

As shown in Table~\ref{tab:tab1}, our method outperforms most baselines across VSR, ASR, and AVSR tasks under both clean and noisy conditions. Although underperforming Llama-AVSR in ASR, our approach provides a unified framework effective across all three tasks. The superior VSR performance particularly demonstrates the effectiveness of "reprogramming" WavLM for lipreading with LLMs, consistent with previous studies~\cite{djilali2023lip2vec}.
With 433h labeled data, as shown in Table~\ref{tab:tab2}, our method achieves significant accuracy improvements. Compared to previous studies using comparable audio-visual data, our approach demonstrates superior performance across all tasks under both clean and noisy conditions. With 1759h labeled data, our method achieves WERs of 20.9\%, 0.84\%, and 0.69\% on the clean LRS3 test set, rivaling or surpassing state-of-the-art methods. However, our VSR results still lag behind LP-Conformer~\cite{10446532}, which uses approximately 100k hours of audio-visual data.

Comparing ASR and AVSR performance on our noisy test set reveals that incorporating visual information significantly improves speech recognition robustness. 
This substantial WER reduction demonstrates that our method successfully fuses speech foundation models with visual representations to enhance noise-invariant capabilities. 
The next section provides detailed AVSR results under noisy conditions.

\subsection{AVSR results on noisy test sets}

\begin{figure*}
    \centering
    \includegraphics[width=0.7\textwidth]{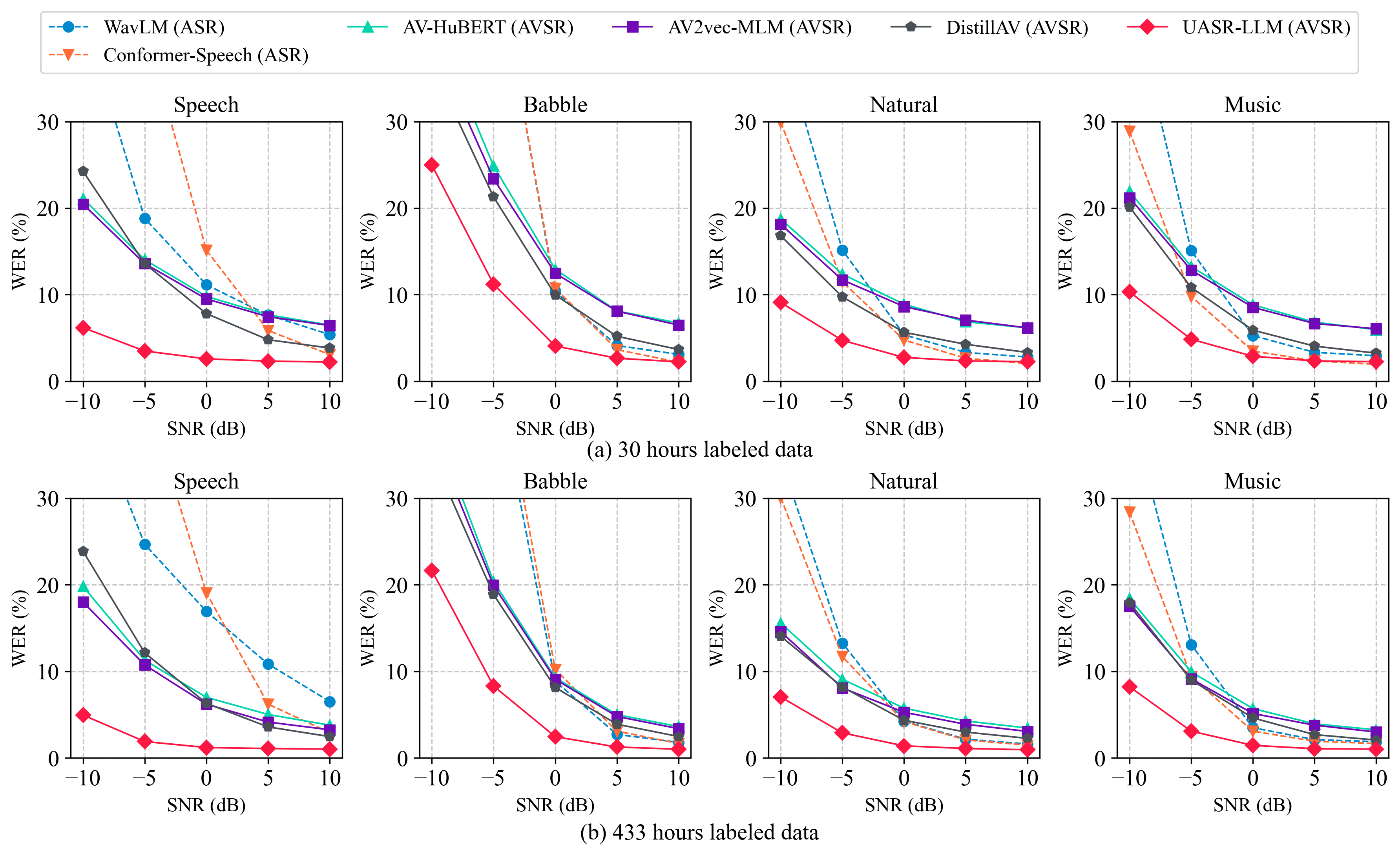}
    \caption{Word error rate (WER) (\%) on our constructed noisy test sets corrupted by various types of noise. }
    %``A'' and ``AV'' represent models using audio and audio-visual inputs respectively.}
    \label{fig:fig4}
\end{figure*}

To evaluate robustness under noisy conditions, we constructed four test sets by mixing the clean audio from the LRS3 test set with different noise types: speech, babble, natural, and music, as described in Section~\ref{sec:implement}. We compared our approach against multiple baselines, including two audio-only SFM models (WavLM and Conformer-Speech) and three AVSR models (AV-HuBERT~\cite{shiavhubert}, AV2vec-MLM~\cite{Zhang2023av2vec}, and DistillAV~\cite{ZHANG2025111432}). All audio-only models were finetuned on the LRS3 dataset. The experiments utilized 433h data for pretraining, with all models adopting the Base configuration. We used either 30h or 433h of LRS3 data for finetuning. 
The experimental results are summarized in Figure~\ref{fig:fig4}.

Our proposed UASR-LLM method (red line) consistently outperforms all baselines across different noise types and SNR levels, with particularly pronounced advantages in severe noise conditions below 0 dB. Among the four noise conditions, babble noise proves most challenging, with all methods exhibiting higher WER at low SNRs, particularly below 0 dB. However, the proposed approach maintains substantially lower error rates than competing methods even in these challenging low-SNR scenarios. The results highlight the effectiveness of incorporating visual information, as audio-only models (WavLM and Conformer-Speech) degrade rapidly in noisy conditions compared to AVSR approaches.

\begin{table*}
    \centering
    \caption{Word error rate (WER) (\%) on the testset of LRS3 with different LLMs and SFMs using the proposed UASR-LLM method.}
    \label{tab:tab3}
    \scalebox{1.0}{
    \begin{tabularx}{0.75\textwidth}{c c c X X X X X}
    \toprule
    Labeled data & LLM type   & SFM type & VSR & ASR & AVSR & ASR (Noisy)  & AVSR (Noisy) \\
     \midrule
     \multicolumn{8}{c}{\emph{DistillAV}~\cite{ZHANG2025111432}} \\
     %\cline{2-8}
     \multirow{7}*{30h} & -- & -- & 36.9 & 2.5 & 2.2 & 44.7 & 9.2 \\
     \cline{2-8}
     \multicolumn{8}{c}{\emph{UASR-LLM (Ours)}} \\
     %\cline{2-8}
     & \multirow{3}*{ChatGLM v3-6B} & Conformer-Speech & 30.0 &  \textbf{1.4} & 1.7 & \textbf{33.6} & 4.8 \\
     & & Whisper & 31.8 & 2.2 & \textbf{1.5} & 39.5 & 4.9 \\
     & & WavLM & \textbf{27.8} & 2.1 & \textbf{1.5} & 40.6 & \textbf{3.7} \\
     % \cline{2-8}
     & \multirow{3}*{Qwen 2.5-7B} & Conformer-Speech & 29.3 & 2.0 & 1.8 & 33.9 & 4.7 \\
     & & Whisper & 32.4 & 2.6 & 1.9 & 34.8 & 5.1 \\
     & & WavLM & 28.4 & 2.4 & 1.9 & 39.8 & 3.8 \\
     \midrule
     \multicolumn{8}{c}{\emph{DistillAV}~\cite{ZHANG2025111432}} \\
     %\cline{2-8}
       \multirow{7}*{433h} & -- & -- & 31.4 & 1.9 & 1.7 & 42.2 & 7.8 \\
     \cline{2-8}
     \multicolumn{8}{c}{\emph{UASR-LLM (Ours)}} \\
     %\cline{2-8}
      & \multirow{3}*{ChatGLM v3-6B} & Conformer-Speech & 24.8 & 
     0.87 & 0.76 & \textbf{31.7} & 3.3 \\
     & & Whisper & 26.1 & 1.3 & 0.81 & 35.1 & 3.9 \\
     & & WavLM & 24.3 & 1.1 & 0.95 & 36.5 & 2.9 \\
     &  \multirow{3}*{Qwen 2.5-7B} & Conformer-Speech & 24.5 & \textbf{0.76} & 0.89 & 32.0 & 3.0 \\
     & & Whisper & 24.6 & 1.2 & \textbf{0.70} & 35.3 & 3.2 \\
     & & WavLM & \textbf{23.9} & 0.86 & 0.78 & 37.0 & \textbf{2.6} \\
     \bottomrule
    \end{tabularx}
    }
\end{table*}

\subsection{Generalization across SFMs and LLMs}

To evaluate the generalization capability of our approach, we conducted experiments with different speech feature models (SFMs) and large language models (LLMs). Specifically, we replaced the WavLM-based SFM with two supervised learning alternatives: Whisper and Conformer-Speech. Additionally, we compared ChatGLM v3-6B with Qwen 2.5-7B as the backbone LLM.
For training, we used 433 hours of data for the visual injection pretraining stage and finetune the models on both 30 hours and 433 hours of the LRS3 dataset. We adopted a pretrained visual encoder (Base version) trained on 1759 hours of data. The experimental results are summarized in Table~\ref{tab:tab3}.

We first analyze the influence of different SFMs using the same LLM and identical amounts of finetuning data. The results show that Conformer-Speech achieves the best ASR performance under both clean and noisy conditions, while WavLM performs best on the VSR task. For AVSR, WavLM achieves superior results on the noisy testset, whereas the clean AVSR results do not exhibit a clear pattern across different SFMs.
When examining the influence of LLMs with the same SFM, we observe that ChatGLM v3-6B slightly outperforms Qwen 2.5-7B when using low-resource data (30 hours) for finetuning. However, this trend reverses with higher-resource data (433 hours), where Qwen 2.5-7B demonstrates better performance.
Despite performance variations across different SFM and LLM combinations, our UASR-LLM consistently achieves significant improvements in recognition accuracy compared to baselines, demonstrating the strong generalization ability of our proposed method.

\subsection{Ablation studies}

\begin{table*}
    \centering
     \caption{WER (\%) results on LRS3 test set for ablation study of visual injection pretraining (VIPT) data amount. "0h" indicates omitting the pretraining stage.}
    \label{tab:tab4}
    \scalebox{1.0}{
    \begin{tabularx}{0.7\textwidth}{c c X X X X X}
    \toprule
   Labeled data  & VIPT data & VSR & ASR & AVSR & ASR (Noisy)  & AVSR (Noisy) \\
   \midrule
   \multirow{3}*{30h} & 0h & 30.1 & \textbf{2.2} & 1.9 & 39.5 & 6.5 \\
   & 433h & 28.4 & 2.4 & 1.9 & 39.8 & 3.8 \\
   & 1759h (Ours) & \textbf{27.3} & 2.3 & \textbf{1.8} & \textbf{38.3} & \textbf{3.5} \\
   \midrule
   \multirow{3}*{433h} & 0h & 26.2 & \textbf{0.78} & 0.78 & \textbf{36.8} & 5.6 \\
   & 433h & 23.9 & 0.86 & 0.78 & 37.0 & 2.6 \\
   & 1759h (Ours) &  \textbf{23.4} & 0.91 & \textbf{0.73} & 37.8 & \textbf{2.4} \\
     \bottomrule
    \end{tabularx}
    }
   
\end{table*}

For our ablation studies, we employed the Base version of the visual encoder pretrained on 1759h of audio-visual data. We used either 30h or 433h of LRS3 data for finetuning. Our ablation studies examined several key design components of our method, including visual injection pretraining, parameter sharing for USR, and the integration of LLM. 

\subsubsection{Visual injection pretraining}
Table~\ref{tab:tab4} summarizes the experimental results of our visual injection pretraining ablation study. The results demonstrate that omitting the proposed visual injection pretraining stage (\textbf{0h} for VIPT data) leads to degraded performance in both VSR and AVSR tasks. Furthermore, increasing the amount of pretraining data during the visual injection pretraining stage yields additional improvements in VSR and AVSR performance. These superior results for VSR and AVSR tasks validate the effectiveness of our pretraining stage in processing visual information.
Interestingly, for the ASR task, the model without the pretraining stage achieves the best results. Since we froze all SFM parameters throughout training, the audio processing capability of SFM itself remained unaffected by the visual injection pretraining stage. This phenomenon may be attributed to our parameter sharing strategy for USR in LLMs, where enhanced AVSR and VSR performance appears to cause a slight degradation in ASR performance—a trade-off we explore in the following section.

\begin{table*}
    \centering
     \caption{WER (\%) results on LRS3 test set for ablation study of parameter sharing and LLM integration. }
    \label{tab:tab5}
    \scalebox{1.0}{
    \begin{tabularx}{0.75\textwidth}{c l  X X X X X}
    \toprule
   Labeled data & Method & VSR & ASR & AVSR & ASR (Noisy)  & AVSR (Noisy) \\
   \midrule
   \multirow{2}*{30h} & UASR-LLM (Ours) & \textbf{28.4} & 2.4 & 1.9 & 39.8 & 3.8  \\
   &  ~~~w/o shared parameters & 30.7 & \textbf{2.2} & \textbf{1.8} & \textbf{38.6} &  \textbf{3.6}  \\
   & ~~~w/o LLM & 32.0 & 5.3 & 2.8 & 40.1 & 6.6 \\
   \midrule
    \multirow{2}*{433h} & UASR-LLM (Ours) & \textbf{23.9} & 0.86 & \textbf{0.78} & 37.0 & \textbf{2.6} \\
   &  ~~~w/o shared parameters & 24.0 & \textbf{0.76} & 0.91 & \textbf{32.8} & 2.9 \\  
   & ~~~w/o LLM & 28.0 & 2.6 & 1.7 & 36.8 & 4.9 \\
   \bottomrule
    \end{tabularx}
    }
   
\end{table*}

\subsubsection{Parameter sharing for USR}
This section compares our parameter sharing approach for realizing USR in a single model with the traditional method of training separate models for each task. The results (\textbf{w/o shared parameters}) are presented in Table~\ref{tab:tab5}. The table shows that training separate models leads to better ASR performance but worse VSR performance. Overall, the proposed USR method achieves comparable performance to the traditional approach that optimizes individual models for each task. These results demonstrate the effectiveness of our method for realizing USR, significantly reducing both training and deployment resources while maintaining strong performance across all tasks.

\subsubsection{Integration of the LLM}
This section compares our proposed method based on a decoder-only LLM with the traditional approach using a 6-layer Transformer-based decoder with cross-attention. The results (\textbf{w/o LLM}) are shown in Table~\ref{tab:tab5}. The table demonstrates that our proposed method employing LLMs significantly outperforms the traditional Transformer decoder. This validates that the strong contextual and language modeling capabilities of LLMs, learned from large text corpora, can be effectively transferred to enhance speech recognition performance, which has also been demonstrated in previous studies~\cite{10889251}.

\section{Conclusion}

In this work, we propose UASR-LLM, a novel approach that adapts speech foundation models (SFMs) for unified multimodal speech recognition, encompassing ASR, VSR, and AVSR within a single framework using a decoder-only LLM for text prediction.
Our method augments the SFM with a visual encoder and inserts visual injection modules into each SFM block to fuse visual representations with audio features. We introduce a visual injection pretraining stage where corrupted audio-visual inputs are processed to predict clean audio representations, followed by speech recognition finetuning with modality dropout. The approach leverages the complementary strengths of SFMs pretrained on large-scale audio corpora and LLMs pretrained on extensive text data.
Experimental results demonstrated the effectiveness of UASR-LLM, achieving superior performance compared to state-of-the-art baselines across ASR, VSR, and AVSR tasks under both clean and noisy conditions. Our method significantly improved recognition robustness across varying signal-to-noise ratios and noise types. Experiments with different SFM and LLM combinations show consistent improvements over baselines, demonstrating the generalization capability of our framework. Ablation studies validated the importance of visual injection pretraining, parameter sharing, and LLM integration. In the future study, we plan to focus on developing more
efficient approaches such as dynamic pooling of adjacent audio-visual representations to reduce input tokens or exploring lightweight LLM architectures. 

%Despite strong performance, our method has several limitations that warrant future investigation. First, while LLM integration leads to substantial speech recognition improvements, it also requires significant computational resources and memory footprint. This computational burden may limit practical deployment, suggesting the need for more efficient approaches such as dynamic pooling of adjacent audio-visual representations to reduce input tokens or exploring lightweight LLM architectures. Second, our method still lags behind previous lipreading studies that employ much larger audio-visual datasets, indicating that the combination of SFMs and LLMs cannot fully compensate for the shortage of large-scale audio-visual training corpora. This limitation fundamentally affects the visual representation extraction capacity of the visual encoder. We anticipate that employing visual encoders pretrained on large-scale visual datasets could further improve our results. Future work will focus on addressing these challenges to enhance both computational efficiency and performance of unified speech recognition systems.

\section{Acknowledgements}
During the preparation of this manuscript, the authors used iFLYTEK’s Xinghuo\footnote{https://xinghuo.xfyun.cn}
 for language editing and readability improvement. After using this tool, the authors carefully reviewed and revised the content as necessary and take full responsibility for the accuracy and integrity of the final manuscript.

%\section{References}

\bibliographystyle{IEEEtran}

\bibliography{ref}

\begin{thebibliography}{10}
\providecommand{\url}[1]{#1}
\csname url@samestyle\endcsname
\providecommand{\newblock}{\relax}
\providecommand{\bibinfo}[2]{#2}
\providecommand{\BIBentrySTDinterwordspacing}{\spaceskip=0pt\relax}
\providecommand{\BIBentryALTinterwordstretchfactor}{4}
\providecommand{\BIBentryALTinterwordspacing}{\spaceskip=\fontdimen2\font plus
\BIBentryALTinterwordstretchfactor\fontdimen3\font minus
  \fontdimen4\font\relax}
\providecommand{\BIBforeignlanguage}[2]{{%
\expandafter\ifx\csname l@#1\endcsname\relax
\typeout{** WARNING: IEEEtran.bst: No hyphenation pattern has been}%
\typeout{** loaded for the language `#1'. Using the pattern for}%
\typeout{** the default language instead.}%
\else
\language=\csname l@#1\endcsname
\fi
#2}}
\providecommand{\BIBdecl}{\relax}
\BIBdecl

\bibitem{9174746}
Y.~Lin, D.~Guo, J.~Zhang, Z.~Chen, and B.~Yang, ``A unified framework for
  multilingual speech recognition in air traffic control systems,'' \emph{IEEE
  Transactions on Neural Networks and Learning Systems}, vol.~32, no.~8, pp.
  3608--3620, 2021.

\bibitem{11021228}
D.~Guo, S.~Zhang, J.~Zhang, B.~Yang, and Y.~Lin, ``Exploring contextual
  knowledge-enhanced speech recognition in air traffic control communication: A
  comparative study,'' \emph{IEEE Transactions on Neural Networks and Learning
  Systems}, vol.~36, no.~9, pp. 16\,085--16\,099, 2025.

\bibitem{JMLR:v25:23-1318}
V.~Pratap, A.~Tjandra, B.~Shi, P.~Tomasello, A.~Babu, S.~Kundu, A.~Elkahky,
  Z.~Ni, A.~Vyas, M.~Fazel-Zarandi, A.~Baevski, Y.~Adi, X.~Zhang, W.-N. Hsu,
  A.~Conneau, and M.~Auli, ``Scaling speech technology to 1,000+ languages,''
  \emph{Journal of Machine Learning Research}, vol.~25, no.~97, pp. 1--52,
  2024.

\bibitem{chen2022wavlm}
S.~Chen, C.~Wang, Z.~Chen, Y.~Wu, S.~Liu, Z.~Chen, J.~Li, N.~Kanda,
  T.~Yoshioka, X.~Xiao \emph{et~al.}, ``{WavLM}: Large-scale self-supervised
  pre-training for full stack speech processing,'' \emph{IEEE Journal of
  Selected Topics in Signal Processing}, vol.~16, no.~6, pp. 1505--1518, 2022.

\bibitem{radford2023robust}
A.~Radford, J.~W. Kim, T.~Xu, G.~Brockman, C.~McLeavey, and I.~Sutskever,
  ``Robust speech recognition via large-scale weak supervision,'' in
  \emph{Proceedings of the International Conference on Machine Learning}.\hskip
  1em plus 0.5em minus 0.4em\relax PMLR, 2023, pp. 28\,492--28\,518.

\bibitem{9837888}
L.~Qu, C.~Weber, and S.~Wermter, ``Lipsound2: Self-supervised pre-training for
  lip-to-speech reconstruction and lip reading,'' \emph{IEEE Transactions on
  Neural Networks and Learning Systems}, vol.~35, no.~2, pp. 2772--2782, 2024.

\bibitem{10682067}
G.~Tan, Z.~Wan, Y.~Wang, Y.~Cao, and Z.-J. Zha, ``Tackling event-based
  lip-reading by exploring multigrained spatiotemporal clues,'' \emph{IEEE
  Transactions on Neural Networks and Learning Systems}, vol.~36, no.~5, pp.
  8279--8291, 2025.

\bibitem{ZHANG2025126741}
J.-X. Zhang, T.~Mao, L.~Guo, J.~Li, and L.~Zhang, ``Target speaker lipreading
  by audio–visual self-distillation pretraining and speaker adaptation,''
  \emph{Expert Systems with Applications}, vol. 272, p. 126741, 2025.

\bibitem{9755926}
Q.~Song, B.~Sun, and S.~Li, ``Multimodal sparse {T}ransformer network for
  audio-visual speech recognition,'' \emph{IEEE Transactions on Neural Networks
  and Learning Systems}, vol.~34, no.~12, pp. 10\,028--10\,038, 2023.

\bibitem{afouras2018deep}
T.~Afouras, J.~S. Chung, A.~Senior, O.~Vinyals, and A.~Zisserman, ``Deep
  audio-visual speech recognition,'' \emph{IEEE Transactions on Pattern
  Analysis and Machine Intelligence}, vol.~44, no.~12, pp. 8717--8727, 2022.

\bibitem{hong2022visual}
J.~Hong, M.~Kim, D.~Yoo, and Y.~Ro, ``Visual context-driven audio feature
  enhancement for robust end-to-end audio-visual speech recognition,'' in
  \emph{Proceedings of the Annual Conference of the International Speech
  Communication Association}, 2022, pp. 2838--2842.

\bibitem{rouditchenko24_interspeech}
A.~Rouditchenko, Y.~Gong, S.~Thomas, L.~Karlinsky, H.~Kuehne, R.~Feris, and
  J.~Glass, ``{Whisper-Flamingo}: Integrating visual features into whisper for
  audio-visual speech recognition and translation,'' in \emph{Proceedings of
  the Annual Conference of the International Speech Communication Association},
  2024, pp. 2420--2424.

\bibitem{petridis2018audio}
S.~Petridis, T.~Stafylakis, P.~Ma, G.~Tzimiropoulos, and M.~Pantic,
  ``Audio-visual speech recognition with a hybrid {CTC}/{Attention}
  architecture,'' in \emph{Proceedings of the Spoken Language Technology
  Workshop}.\hskip 1em plus 0.5em minus 0.4em\relax IEEE, 2018, pp. 513--520.

\bibitem{ma2021end}
P.~Ma, S.~Petridis, and M.~Pantic, ``End-to-end audio-visual speech recognition
  with conformers,'' in \emph{Proceedings of the International Conference on
  Acoustics, Speech and Signal Processing}.\hskip 1em plus 0.5em minus
  0.4em\relax IEEE, 2021, pp. 7613--7617.

\bibitem{shiavhubert}
B.~Shi, W.-N. Hsu, K.~Lakhotia, and A.~Mohamed, ``Learning audio-visual speech
  representation by masked multimodal cluster prediction,'' in
  \emph{Proceedings of the International Conference on Learning
  Representations}, 2022, pp. 1--12.

\bibitem{haliassos2024braven}
A.~Haliassos, A.~Zinonos, R.~Mira, S.~Petridis, and M.~Pantic, ``{BRAVE}n:
  Improving self-supervised pre-training for visual and auditory speech
  recognition,'' in \emph{Proceedings of the IEEE International Conference on
  Acoustics, Speech and Signal Processing}.\hskip 1em plus 0.5em minus
  0.4em\relax IEEE, 2024, pp. 11\,431--11\,435.

\bibitem{shi22_interspeech}
{Bowen Shi and Wei-Ning Hsu and Abdelrahman Mohamed}, ``{Robust Self-Supervised
  Audio-Visual Speech Recognition},'' in \emph{Proceedings of the Annual
  Conference of the International Speech Communication Association}, {2022},
  pp. {2118--2122}.

\bibitem{haliassos2022jointly}
A.~Haliassos, P.~Ma, R.~Mira, S.~Petridis, and M.~Pantic, ``Jointly learning
  visual and auditory speech representations from raw data,'' in
  \emph{Proceedings of the International Conference on Learning
  Representations}, 2023, pp. 1--15.

\bibitem{haliassos2024unified}
A.~Haliassos, R.~Mira, H.~Chen, Z.~Landgraf, S.~Petridis, and M.~Pantic,
  ``Unified speech recognition: A single model for auditory, visual, and
  audiovisual inputs,'' in \emph{Proceedings of Annual Conference on Neural
  Information Processing Systems}, 2024.

\bibitem{djilali2023lip2vec}
Y.~A.~D. Djilali, S.~Narayan, H.~Boussaid, E.~Almazrouei, and M.~Debbah,
  ``{Lip2Vec}: Efficient and robust visual speech recognition via
  latent-to-latent visual to audio representation mapping,'' in
  \emph{Proceedings of the IEEE/CVF International Conference on Computer
  Vision}, 2023, pp. 13\,790--13\,801.

\bibitem{prajwal24_interspeech}
K.~R. Prajwal, T.~Afouras, and A.~Zisserman, ``Speech recognition models are
  strong lip-readers,'' in \emph{Proceedings of the Annual Conference of the
  International Speech Communication Association}, 2024, pp. 2425--2429.

\bibitem{han-etal-2024-xlavs}
H.~Han, M.~Anwar, J.~Pino, W.-N. Hsu, M.~Carpuat, B.~Shi, and C.~Wang,
  ``{XLAVS}-{R}: Cross-lingual audio-visual speech representation learning for
  noise-robust speech perception,'' in \emph{Proceedings of the Annual Meeting
  of the Association for Computational Linguistics}, L.-W. Ku, A.~Martins, and
  V.~Srikumar, Eds.\hskip 1em plus 0.5em minus 0.4em\relax Bangkok, Thailand:
  Association for Computational Linguistics, Aug. 2024, pp. 12\,896--12\,911.

\bibitem{ZHANG2025111432}
J.-X. Zhang, G.~Wan, J.~Gao, and Z.-H. Ling, ``Audio-visual representation
  learning via knowledge distillation from speech foundation models,''
  \emph{Pattern Recognition}, vol. 162, p. 111432, 2025.

\bibitem{baevski2020wav2vec}
A.~Baevski, Y.~Zhou, A.~Mohamed, and M.~Auli, ``{Wav2vec 2.0}: A framework for
  self-supervised learning of speech representations,'' \emph{Proceedings of
  the Advances in Neural Information Processing Systems}, vol.~33, pp.
  12\,449--12\,460, 2020.

\bibitem{10999072}
A.~Rouditchenko, S.~Thomas, H.~Kuehne, R.~Feris, and J.~Glass,
  ``mwhisper-flamingo for multilingual audio-visual noise-robust speech
  recognition,'' \emph{IEEE Signal Processing Letters}, vol.~32, pp.
  2144--2148, 2025.

\bibitem{hu2022lora}
E.~J. Hu, Y.~Shen, P.~Wallis, Z.~Allen-Zhu, Y.~Li, S.~Wang, L.~Wang, and
  W.~Chen, ``Lo{RA}: Low-rank adaptation of large language models,'' in
  \emph{Proceedings of the International Conference on Learning
  Representations}, 2022, pp. 1--13.

\bibitem{hsu2021hubert}
W.-N. Hsu, B.~Bolte, Y.-H.~H. Tsai, K.~Lakhotia, R.~Salakhutdinov, and
  A.~Mohamed, ``{HuBERT}: Self-supervised speech representation learning by
  masked prediction of hidden units,'' \emph{IEEE/ACM Transactions on Audio,
  Speech, and Language Processing}, vol.~29, pp. 3451--3460, 2021.

\bibitem{baevski2022data2vec}
A.~Baevski, W.-N. Hsu, Q.~Xu, A.~Babu, J.~Gu, and M.~Auli, ``Data2vec: A
  general framework for self-supervised learning in speech, vision and
  language,'' in \emph{Proceedings of the International Conference on Machine
  Learning}.\hskip 1em plus 0.5em minus 0.4em\relax PMLR, 2022, pp. 1298--1312.

\bibitem{peng25c_interspeech}
Y.~Peng, M.~Shakeel, Y.~Sudo, W.~Chen, J.~Tian, C.-J. Lin, and S.~Watanabe,
  ``Owsm v4: Improving open whisper-style speech models via data scaling and
  cleaning,'' in \emph{Proceedings of the Annual Conference of the
  International Speech Communication Association}, 2025, pp. 2225--2229.

\bibitem{Chung_2017_CVPR}
J.~Son~Chung, A.~Senior, O.~Vinyals, and A.~Zisserman, ``Lip reading sentences
  in the wild,'' in \emph{Proceedings of the IEEE/CVF Conference on Computer
  Vision and Pattern Recognition}, July 2017, pp. 6447--6456.

\bibitem{10096889}
P.~Ma, A.~Haliassos, A.~Fernandez-Lopez, H.~Chen, S.~Petridis, and M.~Pantic,
  ``Auto-{AVSR}: Audio-visual speech recognition with automatic labels,'' in
  \emph{IEEE International Conference on Acoustics, Speech and Signal
  Processing}, 2023, pp. 1--5.

\bibitem{sterpu2020teach}
G.~Sterpu, C.~Saam, and N.~Harte, ``How to teach {DNN}s to pay attention to the
  visual modality in speech recognition,'' \emph{IEEE/ACM Transactions on
  Audio, Speech, and Language Processing}, vol.~28, pp. 1052--1064, 2020.

\bibitem{ma2022training}
P.~Ma, Y.~Wang, S.~Petridis, J.~Shen, and M.~Pantic, ``Training strategies for
  improved lip-reading,'' in \emph{Proceedings of the International Conference
  on Acoustics, Speech and Signal Processing}.\hskip 1em plus 0.5em minus
  0.4em\relax IEEE, 2022, pp. 8472--8476.

\bibitem{afouras2020asr}
T.~Afouras, J.~S. Chung, and A.~Zisserman, ``{ASR} is all you need: Cross-modal
  distillation for lip reading,'' in \emph{Proceedings of the International
  Conference on Acoustics, Speech and Signal Processing}.\hskip 1em plus 0.5em
  minus 0.4em\relax IEEE, 2020, pp. 2143--2147.

\bibitem{hsu2022u}
W.-N. Hsu and B.~Shi, ``{u-HuBERT}: Unified mixed-modal speech pretraining and
  zero-shot transfer to unlabeled modality,'' \emph{Proceedings of the Advances
  in Neural Information Processing Systems}, vol.~35, pp. 21\,157--21\,170,
  2022.

\bibitem{Zhang2023av2vec}
J.-X. Zhang, G.~Wan, Z.-H. Ling, J.~Pan, J.~Gao, and C.~Liu, ``{Self-supervised
  audio-visual speech representations learning by multimodal
  self-distillation},'' in \emph{Proceedings of the International Conference on
  Acoustics, Speech and Signal Processing}, 2023, pp. 1--5.

\bibitem{Lian2023avdata2vec}
J.~Lian, A.~Baevski, W.-N. Hsu, and M.~Auli, ``{AV-data2vec: Self-supervised
  learning of audio-visual speech representations with contextualized target
  representations},'' in \emph{Proceedings of the IEEE Automatic Speech
  Recognition and Understanding Workshop}, 2023, pp. 1--8.

\bibitem{zhang-etal-2023-speechgpt}
D.~Zhang, S.~Li, X.~Zhang, J.~Zhan, P.~Wang, Y.~Zhou, and X.~Qiu,
  ``{S}peech{GPT}: Empowering large language models with intrinsic cross-modal
  conversational abilities,'' in \emph{Proceedings of the Conference on
  Empirical Methods in Natural Language Processing}, H.~Bouamor, J.~Pino, and
  K.~Bali, Eds.\hskip 1em plus 0.5em minus 0.4em\relax Singapore: Association
  for Computational Linguistics, Dec. 2023, pp. 15\,757--15\,773.

\bibitem{Qwen2.5-Omni}
J.~Xu, Z.~Guo, J.~He, H.~Hu \emph{et~al.}, ``Qwen2.5-omni technical report,''
  \emph{arXiv preprint arXiv:2503.20215}, 2025.

\bibitem{10389703}
M.~Wang, W.~Han, I.~Shafran, Z.~Wu, C.-C. Chiu, Y.~Cao, N.~Chen, Y.~Zhang,
  H.~Soltau, P.~K. Rubenstein, L.~Zilka, D.~Yu, G.~Pundak, N.~Siddhartha,
  J.~Schalkwyk, and Y.~Wu, ``{SLM}: Bridge the thin gap between speech and text
  foundation models,'' in \emph{Proceedings of the IEEE Automatic Speech
  Recognition and Understanding Workshop}, 2023, pp. 1--8.

\bibitem{10447605}
Y.~Fathullah, C.~Wu, E.~Lakomkin, J.~Jia, Y.~Shangguan, K.~Li, J.~Guo,
  W.~Xiong, J.~Mahadeokar, O.~Kalinli, C.~Fuegen, and M.~Seltzer, ``Prompting
  large language models with speech recognition abilities,'' in
  \emph{Proceedings of the IEEE International Conference on Acoustics, Speech
  and Signal Processing}, 2024, pp. 13\,351--13\,355.

\bibitem{10445874}
W.~Yu, C.~Tang, G.~Sun, X.~Chen, T.~Tan, W.~Li, L.~Lu, Z.~Ma, and C.~Zhang,
  ``Connecting speech encoder and large language model for {ASR},'' in
  \emph{Proceedings of the IEEE International Conference on Acoustics, Speech
  and Signal Processing}, 2024, pp. 12\,637--12\,641.

\bibitem{xu24d_interspeech}
Y.~Xu, S.-X. Zhang, J.~Yu, Z.~Wu, and D.~Yu, ``Comparing discrete and
  continuous space llms for speech recognition,'' in \emph{Proceedings of the
  Annual Conference of the International Speech Communication Association},
  2024, pp. 2509--2513.

\bibitem{yeo-etal-2024-visual}
J.~Yeo, S.~Han, M.~Kim, and Y.~M. Ro, ``Where visual speech meets language:
  {VSP}-{LLM} framework for efficient and context-aware visual speech
  processing,'' in \emph{Proceedings of the Conference on Empirical Methods in
  Natural Language Processing}, 2024, pp. 11\,391--11\,406.

\bibitem{yang24f_interspeech}
G.~Yang, Z.~Ma, F.~Yu, Z.~Gao, S.~Zhang, and X.~Chen, ``Mala-{ASR}:
  Multimedia-assisted {LLM}-based {ASR},'' in \emph{Proceedings of the Annual
  Conference of the International Speech Communication Association}, 2024, pp.
  2405--2409.

\bibitem{10889251}
U.~Cappellazzo, M.~Kim, H.~Chen, P.~Ma, S.~Petridis, D.~Falavigna, A.~Brutti,
  and M.~Pantic, ``Large language models are strong audio-visual speech
  recognition learners,'' in \emph{Proceedings of the IEEE International
  Conference on Acoustics, Speech and Signal Processing}, 2025, pp. 1--5.

\bibitem{yeo2025mms}
J.~H. Yeo, H.~Rha, S.~J. Park, and Y.~M. Ro, ``{MMS}-{LL}a{MA}: Efficient
  llm-based audio-visual speech recognition with minimal multimodal speech
  tokens,'' in \emph{Proceedings of the Annual Meeting of the Association for
  Computational Linguistics}, 2025, pp. 20\,724--20\,735.

\bibitem{shaw-etal-2018-self}
P.~Shaw, J.~Uszkoreit, and A.~Vaswani, ``Self-attention with relative position
  representations,'' in \emph{Proceedings of the Conference of the North
  {A}merican Chapter of the Association for Computational Linguistics},
  M.~Walker, H.~Ji, and A.~Stent, Eds.\hskip 1em plus 0.5em minus 0.4em\relax
  New Orleans, Louisiana: Association for Computational Linguistics, Jun. 2018,
  pp. 464--468.

\bibitem{Zhang2022face}
J.-X. Zhang, G.~Wan, and J.~Pan, ``{Is lip region-of-interest sufficient for
  lipreading?}'' in \emph{Proceedings of the International Conference on
  Multimodal Interaction}, 2022, pp. 368–--372.

\bibitem{DBLP:journals/corr/abs-2201-01763}
B.~Shi, W.~Hsu, and A.~Mohamed, ``Robust self-supervised audio-visual speech
  recognition,'' in \emph{Proceedings of the Annual Conference of the
  International Speech Communication Association}, 2022, pp. 2118--2122.

\bibitem{zhu2023vatlm}
Q.~Zhu, L.~Zhou, Z.~Zhang, S.~Liu, B.~Jiao, J.~Zhang, L.~Dai, D.~Jiang, J.~Li,
  and F.~Wei, ``{VATLM}: Visual-audio-text pre-training with unified masked
  prediction for speech representation learning,'' \emph{IEEE Transactions on
  Multimedia}, vol.~26, pp. 1055--1064, 2024.

\bibitem{makino2019recurrent}
T.~Makino, H.~Liao, Y.~Assael, B.~Shillingford, B.~Garcia, O.~Braga, and
  O.~Siohan, ``Recurrent neural network transducer for audio-visual speech
  recognition,'' in \emph{Proceedings of the Automatic Speech Recognition and
  Understanding Workshop}.\hskip 1em plus 0.5em minus 0.4em\relax IEEE, 2019,
  pp. 905--912.

\bibitem{10446532}
O.~Chang, H.~Liao, D.~Serdyuk, A.~Shah, and O.~Siohan, ``Conformer is all you
  need for visual speech recognition,'' in \emph{Proceedings of the
  International Conference on Acoustics, Speech and Signal Processing}, 2024,
  pp. 10\,136--10\,140.

\end{thebibliography}

\end{document}